\documentclass[acmlarge,manuscript]{acmart}

\AtBeginDocument{%
  }

\setcopyright{acmlicensed}
\copyrightyear{2018}
\acmYear{2018}
\acmDOI{XXXXXXX.XXXXXXX}

\acmJournal{IMWUT}
\acmVolume{37}
\acmNumber{4}
\acmArticle{111}
\acmMonth{8}

\usepackage{tabularx}
\usepackage{booktabs}
\usepackage{siunitx} 
\usepackage{makecell}
\usepackage{multirow}   
\usepackage{svg} 
\usepackage{graphicx}
\usepackage{pifont}
\usepackage{soul, xcolor}
\usepackage{longtable}
\usepackage{listings}
\usepackage{xcolor}

\lstdefinestyle{cli}{
  basicstyle=\ttfamily\small,
  columns=fullflexible,
  breaklines=true,
  breakatwhitespace=true,
  frame=single,
  framerule=0.3pt,
  rulecolor=\color{black!25},
  xleftmargin=0pt,
  xrightmargin=0pt,
  framexleftmargin=0pt,
  framexrightmargin=0pt,
  aboveskip=0.6em,
  belowskip=0.6em,
  showstringspaces=false,
  keepspaces=true
}

\begin{document}
\title[Do LLMs Need to See Everything?]{Do LLMs Need to See Everything? A Benchmark and Study of Failures in LLM-driven Smartphone Automation using Screentext vs. Screenshots}

\author{Shiquan Zhang}
\email{shiquan.zhang@student.unimelb.edu.au}
\orcid{0000-0003-3747-0842}
\affiliation{%
  \department{School of Computing and \\Information Systems}
  \institution{University of Melbourne}
  \city{Melbourne}
  \country{Australia}
}

\author{Tianyi Zhang}
\email{t.zhang59@student.unimelb.edu.au}
\orcid{0000-0002-0778-8844}
\affiliation{%
  \department{School of Computing and \\Information Systems}
  \institution{University of Melbourne}
  \city{Melbourne}
  \country{Australia}
}

\author{Le Fang} 
\email{le.fang2@unimelb.edu.au}
\orcid{0009-0004-2591-7584}
\affiliation{%
  \department{School of Computing and \\Information Systems}
  \institution{University of Melbourne}
  \city{Melbourne}
  \country{Australia}
}

\author{Simon D'Alfonso}
\email{dalfonso@unimelb.edu.au}
\orcid{0000-0001-7407-8730}
\affiliation{%
  \department{School of Computing and \\Information Systems}
  \institution{University of Melbourne}
  \city{Melbourne}
  \country{Australia}
}

\author{Hong Jia}
\email{hong.jia@auckland.ac.nz}
\orcid{0000-0002-6047-4158}
\affiliation{%
  \department{School of Computer Science}
  \institution{University of Auckland}
  \city{Auckland}
  \country{New Zealand}
}

\author{Vassilis Kostakos}
\email{vassilis.kostakos@unimelb.edu.au}
\orcid{0000-0003-2804-6038}
\affiliation{%
  \department{School of Computing and \\Information Systems}
  \institution{University of Melbourne}
  \city{Melbourne}
  \country{Australia}
}

\renewcommand{\shortauthors}{Zhang et al.}
\begin{abstract}
    With the rapid advancement of large language models (LLMs), mobile agents have emerged as promising tools for phone automation, simulating human interactions on screens to accomplish complex tasks. However, these agents often suffer from low accuracy, misinterpretation of user instructions, and failure on challenging tasks, with limited prior work examining why and where they fail. To address this, we introduce DailyDroid, a benchmark of 75 tasks in five scenarios across 25 Android apps, spanning three difficulty levels to mimic everyday smartphone use. We evaluate it using text-only and multimodal (text + screenshot) inputs on GPT-4o and o4-mini across 300 trials, revealing comparable performance with multimodal inputs yielding marginally higher success rates. Through in-depth failure analysis, we compile a handbook of common failures. Our findings reveal critical issues in UI accessibility, input modalities, and LLM/app design, offering implications for future mobile agents, applications, and UI development.
\end{abstract}


\received{20 February 2007}
\received[revised]{12 March 2009}
\received[accepted]{5 June 2009}


\maketitle

\section{Introduction}
Mobile agents understand natural language instructions and can complete mobile tasks that are often repetitive and complex. Examples include information retrieval such as news, weather, and transportation through voice interaction with Siri \citep{siri}; personalized assistance that makes recommendations based on user preferences and context \citep{jones2024designing}; and intent interpretation for tasks like booking a restaurant \citep{hong2024cogagent}.

Powered by recent breakthroughs in Large Language Models (LLMs) like ChatGPT \cite{OpenAI2022ChatGPT} and Claude \cite{claude2025}, mobile agents now leverage screenshots as input and use LLMs as the "brain" to analyze, predict, and execute complex tasks \citep{anthropic_claude_2024, openai:computeragent, hong2024cogagent}. Focusing on autonomous systems capable of operating within mobile environments, these agents integrate multimodal perception, planning, action execution, and memory to navigate dynamic interfaces across applications, web browsers, and operating systems \cite{wu2024foundations}. Mobile task automation builds on these capabilities to enable agents to perform tasks by directly interacting with a device's Graphical User Interface (GUI). The primary goal is to automatically execute complex actions based on natural language instructions, allowing for hands-free operation and smoother interaction \citep{VisionTasker2024, wen2024autodroid}. However, most of these agents suffer from low accuracy, misunderstanding of user intent, and premature termination \citep{liu2025llm}, all of which significantly impede the user experience. Surprisingly, little work has investigated exactly why and where these models fail to complete tasks on mobile devices \cite{cemri2025multi, schombs2025conversation, liu2025llm}. At the same time, sharing screenshots of one's phone can feel invasive \citep{reeves2020time}. This raises a natural question: do we really need screenshots as the input modality for mobile agents?

In this paper we first propose DailyDroid, a mobile agent benchmark designed to systematically investigate why and where these agents succeed or fail. We create 75 tasks across five common daily scenario each including five representative apps and three difficulty levels. We then use our benchmark to test two representative LLMs: GPT-4o as a baseline and o4-mini as a more advanced model with reasoning capability. We consider two possible modalities as input to the models: using only "screentext" which consists of the UI tree in Android, and "screenshot" which is representative of how most agents work in recent years, by providing a screenshot along with the screentext.  
The main contributions of this paper are summarized as follows:
\begin{itemize}
    \item We propose and deploy a benchmark to systematically evaluate in-depth the kinds of mistakes that agents make when completing tasks on mobile devices.
    \item We compare input modalities by analyzing collected screentext versus screenshots, and evaluate model performance by contrasting reasoning-oriented LLMs with baseline LLMs.
    \item Our findings provide design implications and recommendations for future mobile apps and devices, LLM agents, and UI development.
\end{itemize}

\section{Related Work}

\subsection{Intelligent Personal Assistants}
The concept of Intelligent Personal Assistants (IPAs), software agents designed to augment individual abilities, complete complex tasks, and potentially satisfy emotional needs \cite{li2024personal}, has a long history. Dating back to the early 1950s, Bell Labs developed “Audrey”, a speech recognition system capable of recognizing digits spoken by a single voice \citep{Sonix2024History,li2024personal}. Then in the 1990s, "Dragon Dictate" \citep{bamberg1990dragon}, a voice-based software, was first created for commercial products and designed to support discrete speech recognition on Microsoft Windows. In the next two decades, advancements in natural language processing led to voice-activated assistants on smartphones like Apple's Siri \citep{siri} (2011), Amazon's Alexa \citep{Alexa} (2014), and Google Assistant \citep{Googleassistant} (2016). In November 2022, OpenAI released ChatGPT \citep{OpenAI2022ChatGPT}, an LLM-powered chatbot that enabled interaction via a simple input-box interface and significantly changed how humans interact with machines. Since then, the field has developed rapidly, with related products emerging in quick succession. In November 2023, Microsoft launched the landmark Microsoft Copilot \citep{Microsoft2023Copilot}, designed to create a range of personal assistants for devices such as computers and smartphones. Early generative agents, such as those demonstrated in \cite{weng2023agent,park2023generative,hong2023metagpt,wang2024survey} established a foundational architecture for simulating human-like behavior using LLMs, planning, memory, execution, and reflection. This approach re-framed agents as contextual and adaptive assistants, paving the way for subsequent breakthroughs in IPAs.

Building on this foundation, recent developments have substantially expanded their practical capabilities in real-world contexts. By the end of 2024, Anthropic released the first frontier AI model to offer computer use in public \cite{anthropic_claude_2024}, which allowed users to direct Claude to use computers the way people do by looking at a screen, moving a cursor, clicking buttons, and typing text. Following closely behind, another computer-using agent, Operator, from OpenAI, powered by GPT-4o's vision capabilities, captured screenshots in a web browser and used a virtual mouse and keyboard to complete actions \cite{openai:computeragent}. Four months later, Butterfly Effect launched Manus \cite{manus2025}, the first autonomous AI agent system designed to collaboratively plan and execute complex tasks across a web browser. Essentially, these systems take input from GUIs, employ LLMs to plan and reason, and execute actions on platforms, such as computers, browsers, and smartphones. In general, recent works have enabled human AI interaction from simple chit-chat to more intelligent and complex task automation. However, automating these complex tasks remains a significant challenge \citep{pan2023autotask, openai:computeragent, kahlon2025agent, liu2025llm}, frequently leading to system failures and degraded performance. Recent research highlights a range of persistent issues, including biases in perception and understanding \citep{gur2024realworldwebagentplanninglong, niu2024screenagent} such as element misalignment and visual confusion—as well as limitations in reasoning and decision-making \citep{ZhangTWW0HTLZ024, MerrillPS24}, for example, difficulties with contextual comprehension and abstract inference. Moreover, the complexity of dynamic web environments \citep{ZhouX0ZLSCOBF0N24}, including dynamic content and network delays, introduces further obstacles. Collectively, these challenges hinder the progress of current IPAs powered by LLMs and agentic frameworks.

Tracing back to the historical development of IPAs, most features/tools supported by various technologies can be traced back to AI agents. An agent, in the context of artificial intelligence, is an entity that perceives its environment through sensors and acts upon that environment through actuators to achieve specific goals \cite{liu2025llm, weng2023agent}. With the advent of LLMs and the rapid growth of the open-source community, interest in IPAs has been reignited. This long-established field is experiencing a resurgence, with developments emerging at an unprecedented pace across both academia and industry \cite{wang2024survey}.

Yet, despite this progress, little is known about why and where these agents fail in everyday smartphone use. To address this gap, we propose and deploy \textit{DailyDroid}, a benchmark of everyday mobile tasks, and conduct a systematic failure analysis of LLM-driven mobile agents, offering insights for designing more reliable and intelligent IPAs.

\subsection{Mobile Task Automation}
\label{mobiletaskautmation}

Automating tasks on mobile user interfaces presents distinct challenges compared to computers or web browsers \citep{wu2024foundations,liu2025llm}. Mobile interactions are characterized by smaller screen real estate, touch-based inputs \citep{Zhang2023AppAgentMA} such as tapping, swiping, and typing, and highly dynamic interfaces \citep{google_fragments_guide} that vary significantly across applications and operating system versions, and resource constraints \citep{hosio2016monetary}.

Traditional automation methods, often relying on brittle predefined scripts or fixed templates, while effective, struggle with their complex and variable UI and dynamic settings when facing issues such as limited generalizability across apps, high maintenance costs, and poor understanding of user intent or visual context \citep{liu2025llm}. These methods can be viewed as early forms of agents, designed to perform specific tasks in a predetermined manner. In recent years, the rapid advancement of mobile technologies and the emergence of LLMs have spurred the development of sophisticated mobile agents capable of interpreting natural language instructions and interacting with mobile GUIs. In this context, mobile agents are autonomous computational entities designed to operate within mobile device environments. Several frameworks, such as AppAgent Series \citep{Zhang2023AppAgentMA, li2024appagent2, jiang2025appagentx}, Mobile-Agent Series \citep{wang2024mobile, wang2024mobile2}, and AutoTask \citep{pan2023autotask}, have been proposed to orchestrate the creation and deployment of these agents. Despite progress, existing frameworks still struggle to remain robust amid the diversity and dynamics of mobile apps \citep{gur2024realworldwebagentplanninglong, niu2024screenagent}, grounding actions to the correct UI elements (especially visually similar or complex ones) \citep{cheng2024seeclick, baechler2024screenai}, handling long multi-step tasks, managing state changes, and recovering from errors \cite{shvo2021appbuddy, pan2023autotask}. 

Focusing on these problems, our work proposes a comprehensive benchmark to reveal why and where these mobile agents fail, and provides design implications for future mobile agent design. 

\subsection{Screen Representation for LLMs}
\label{screencontent}
LLM-driven mobile agents rely on screen representations to perceive and reason about UI states. Prior work uses two main families—UI-tree (textual) representations and screenshot (pixel) representations—as well as hybrid approaches that combine pixels with structure or OCR \cite{liu2025llm}.

\subsubsection{Screentext Representations}
A UI tree is a structured, hierarchical, textual description of user interface elements. Each node corresponds to a component with attributes (e.g., class, visibility, resource IDs). Importantly, the UI tree not only encodes the text and components visible on the screen, but also includes hidden or off-screen elements that may not be directly perceivable. In this paper, we refer to UI-tree representations as \textit{screentext (text-only)}. Common sources include the View Hierarchy (VH) \citep{wang2024mobileagentbench}, Accessibility Services \citep{zhong2025screenaudit, android:a11yservice}, and simplified formats that normalize VH to HTML/XML for efficiency \citep{wang2023enabling, wen2024autodroid}. Examples include transforming UI trees into natural-language descriptions \citep{wen2023droidbot}, mapping VH to HTML \citep{wang2023enabling}, and parsing GUIs into simplified HTML \citep{wen2024autodroid}. Despite their practicality, text-only inputs face well-noted issues: not all on-screen content is exposed via text APIs \citep{song2024visiontasker, shaw2023pixels, cheng2024seeclick}; the extracted text can be verbose or incomplete \citep{deng2023mind2web, teng2024tool, zhang2024enabling}; and there is no standardized simplification parsing pipeline for LLMs \citep{kim2023language, wang2023enabling, pan2023autotask}.

\subsubsection{Screenshot Representations}
A screenshot captures the visual appearance, including layout, styling, text, and images of the screen, exactly as seen by a human user. Unlike UI trees, which require API access and can become unwieldy with complex hierarchies, screenshots offer a more flexible and often more comprehensive representation of the UI. In this paper, we refer to the combination of screenshots and screentext as \textit{multimodal} input. AutoUI \citep{zhang2023you} demonstrates effective GUI control with only screenshots, while ScreenAI \cite{baechler2024screenai} and SeeClick \cite{cheng2024seeclick} show that learning directly from screenshots with richer visual info enhances LLMs' adaptability. Others, CogAgent \cite{hong2024cogagent}, Appagent \cite{Zhang2023AppAgentMA}, MobileGPT \cite{lee2023explore}, leverage screenshots in combination with supplementary information to enhance UI understanding. However, pixel-centric methods suffer from high latency and cost \citep{liu2025llm}, require complex and unreliable annotations \citep{Zhang2023AppAgentMA}, and introduce privacy concerns as screenshots may reveal sensitive user information \citep{reeves2020time, teng2024tool}.

Input representation fundamentally shapes what an agent can perceive and, consequently, its capabilities and results. Our work incorporates this factor by systematically comparing text-only versus multimodal for mobile task automation.

\begin{figure}[h]
  \centering
  \includegraphics[width=\linewidth]{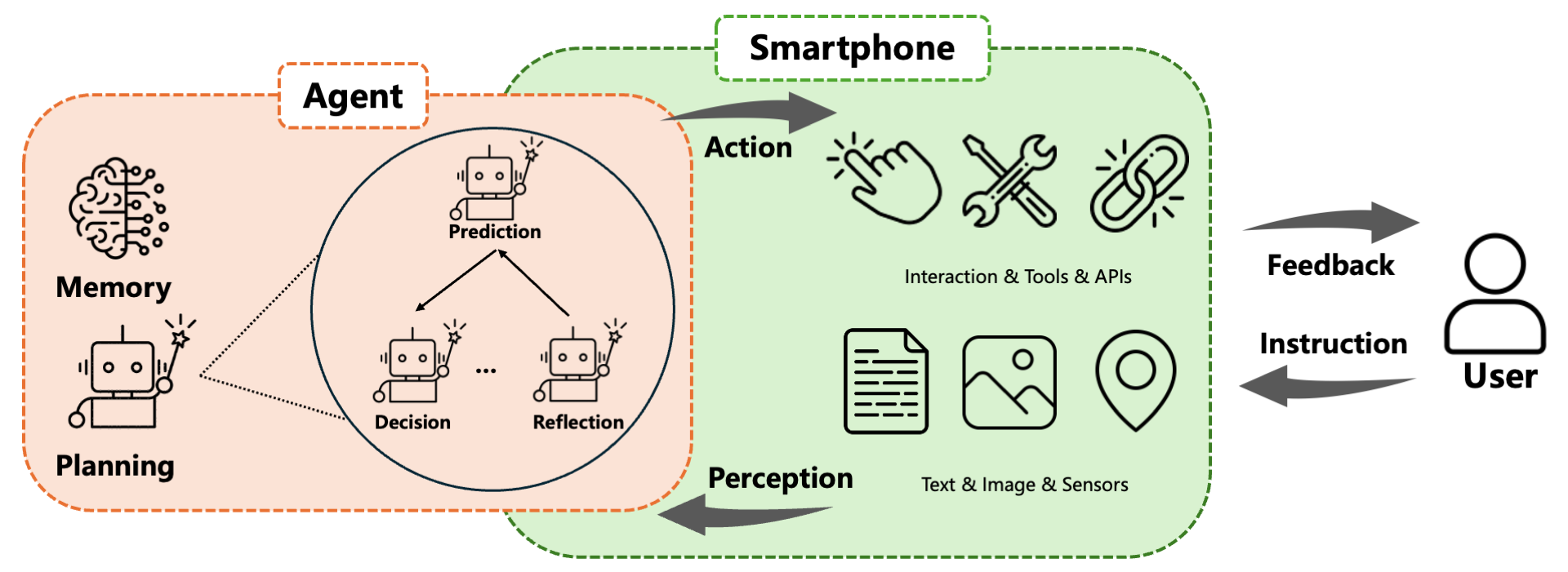}
  \caption{An overview of the Mobile Agent System. It depicts the interaction between the user, the smartphone environment, and the agent, involving Perception, Planning, Action, and Memory modules. Perception module captures the smartphone environment's state, such as text, image, and sensors. Planning module, including sub-modules such as Prediction, Decision, and Reflections, etc., uses this input and knowledge from the Memory module to make decisions, which the Action module executes to interact with the environment.}
  \Description{A block diagram showing the Mobile Agent System. On the left, the Agent includes Memory and Planning, with submodules Prediction, Decision, and Reflection. In the center, the Smartphone block is divided into Perception (text, image, and sensors) and Action (interactions, tools, and APIs). On the right, a User provides instructions to the smartphone and receives feedback. Arrows indicate the flow: User to Smartphone, Smartphone to Agent, and back.}
  \label{mobileagent}
\end{figure}

\section{Method}

\subsection{Preliminaries: Mobile Agent}
\label{mobileagentsection}
Similar to the general LLM-Powered autonomous agent system \cite{weng2023agent}, a mobile agent system involves distinct components for Perception, Planning, Action, and Memory. Together, these components enable agents to perceive, reason, and execute within dynamic mobile environments, adapting their behavior dynamically to improve task efficiency and robustness (Figure \ref{mobileagent}).

\textbf{Perception}. Perception serves as the initial step to capture and interpret the state of the mobile environment. Typically, Perception has two kinds of input: UI information, including UI tree-based and screenshot-based approaches \cite{liu2025llm}, and Phone state information from smartphone sensors \cite{zhang2024aware,jones2024designing}. 


\textbf{Planning}. Planning is the brain of mobile agents, formulating action strategies based on task objectives and dynamic environments. Receiving the input prompts including UI state, historical interactions, and relevant knowledge, LLMs conduct reasoning, evaluate possible actions with reasons, and select the most appropriate ones to achieve the desired outcomes \citep{weng2023agent, liu2025llm}. At a deeper level, certain works \citep{shvo2021appbuddy,wu2022survey,pan2023autotask} extend the role of LLMs in planning by incorporating capabilities for reflection and backtracking.

\textbf{Action}. Action is responsible for executing decisions made by the Planning module. By bridging high-level commands from LLMs with low-level device operations, the agent can issue precise commands or action space \citep{Zhang2023AppAgentMA,xu2024androidlab}, such as tap, scroll, and type, using tools like Android’s UI Automator and Android Debug Bridge.

\textbf{Memory}. Memory enables agents to retain and retrieve existing information to inform the Planning component. Commonly, short-term memory is injected into LLMs through prompt engineering \cite{weng2023agent} and external vector store \cite{Zhang2023AppAgentMA, wen2024autodroid, pan2023autotask}, while long-term memory is incorporated through pre-trained knowledge and fine-tuning \cite{weng2023agent, liu2025llm, liu2025advances}.

\subsection{Benchmark}

\subsubsection{Motivation}
LLM-driven mobile agents are increasingly positioned as a practical interface layer for accomplishing everyday smartphone tasks. To evaluate progress, the community has proposed benchmarks that improve standardization and reproducibility by specifying task suites, environments, and success checks. These resources have standardized evaluation protocols, fostered reproducibility, and driven measurable progress in mobile task automation.

\begin{table}[h]
  \caption{Representative mobile agent benchmarks and the analyses emphasized in their original papers.}
  \Description{The table compares MobileAgentBench, A3, AndroidWorld, and DailyDroid by tasks, app coverage, whether the benchmark targets mainstream third-party apps, whether the original paper reports modality tradeoffs, and whether it provides a reusable diagnostic resource (e.g., failure taxonomy/labels or standardized traces) to support reproducible failure characterization.}
  \label{tab:benchmarks}
  \centering
  \small
  \setlength{\tabcolsep}{4pt}
  \begin{tabular*}{\textwidth}{@{\extracolsep{\fill}}l c c c c c}
    \toprule
    \textbf{Name} &
    \textbf{Tasks} &
    \textbf{\makecell{Apps}} &
    \textbf{\makecell{Mainstream \\ 3P Apps}} &
    \textbf{\makecell{Reported \\ Modality Study}} &
    \textbf{\makecell{Reusable \\ Diagnostic Resource}} \\
    \midrule
    MobileAgentBench \citep{wang2024mobileagentbench} & 100 & 10 & \ding{55} & \ding{55} & \ding{55} \\
    A3 \citep{chai2025a3} & 201 & 21 & \ding{51} & \ding{55} & \ding{55} \\
    AndroidWorld \citep{rawles2024androidworld} & 116 & 20 & \ding{51} & \ding{55} & \ding{55} \\
    \midrule
    \textbf{DailyDroid (Ours)} & \textbf{75} & \textbf{25} & \textbf{\ding{51}} & \textbf{\ding{51}} & \textbf{\ding{51}} \\
    \bottomrule
  \end{tabular*}
\end{table}

Table~\ref{tab:benchmarks} highlights representative benchmarks that have advanced mobile agent evaluation. However, when the goal is to understand why agents fail under everyday smartphone conditions and what those failures imply for human-centered design, existing benchmarks still leave important gaps. Despite their contributions, existing benchmarks reveal three notable gaps that are central to this study:

\begin{enumerate}
    \item \textbf{Everyday failures are under-surfaced by benchmark design choices.} Many benchmarks\citep{Zhang2023AppAgentMA, liu2025llm,rawles2024androidworld,chai2025a3} necessarily prioritize reproducibility and scalable evaluation, for example by constraining app choices, relying on controlled initialization and success checks, or focusing on tasks that are straightforward to verify automatically. These design choices are valuable, but they can under-sample the failure modes that dominate real deployments, such as missing or inconsistent accessibility metadata, dynamic feeds and pop-ups, permission and security prompts, and cross-app dependencies. As a result, common breakdowns that emerge only when agents operate on widely used, fast-evolving consumer apps are not comprehensively exposed, limiting what we can learn about the practical barriers to mobile automation.
    \item \textbf{Modality is also a privacy and ecosystem-acceptance problem, not only an accuracy problem.} From a systems perspective, most mobile GUI agents are screenshot-centric\citep{Zhang2023AppAgentMA, pan2023autotask,liu2025llm} because pixels preserve layout, iconography, and visual cues that are often missing from UI trees. Yet transmitting smartphone screenshots to LLMs can be highly sensitive, as screenshots may contain private messages, sensitive information. This concern is no longer hypothetical: recent industrial attempts to integrate OS-level “phone assistants” have triggered immediate ecosystem pushback. For example, ByteDance’s Doubao phone assistant, deployed on a ZTE Nubia prototype and designed to operate apps via privileged, screen-level access, faced rapid restrictions by widely used apps and was forced to scale back capabilities amid security and privacy concerns \citep{scientificamerican2026}. This motivates a human-centered question that benchmarks rarely operationalize: \emph{when is screenshot access justified, and can agents achieve comparable utility with less invasive representations such as screentext from accessibility APIs?}
    \item \textbf{Limited HCI-centered, repeatable evaluation resources for cumulative knowledge.} The current mobile agent literature is dominated by AI/system-oriented contributions that optimize systems, models, and algorithms, while comparatively fewer works originate from HCI naperspectives~\citep{cemri2025multi, pan2025measuring} that foreground user risk, interaction breakdowns, and design implications. When HCI studies do evaluate agent behavior, the evidence is often generated through one-off, context-specific methods~\citep{schwind2023hci} (e.g., interviews, surveys, lab studies, or bespoke field deployments) that are difficult to replicate, re-run after model updates, or compare across teams. In contrast, a benchmark can enable \emph{cumulative} evaluation by providing shared tasks, controlled conditions, and reusable traces and labels. Our review of existing benchmarks suggests that few explicitly encode HCI-relevant questions (e.g., privacy exposure, accessibility dependence, user-visible failure costs) into repeatable protocols, and few release diagnostic artifacts that allow the community to systematically characterize breakdowns over time. Consequently, the field lacks a standardized, human-centered evaluation substrate that supports both rigorous comparison and actionable implications for mobile UI and agent design.
\end{enumerate}

To address these gaps, we introduce \textit{DailyDroid}, a benchmark purpose-built to complement existing evaluations by emphasizing \textbf{(i) everyday, in-the-wild task friction}, \textbf{(ii) privacy-aware modality tradeoffs}, and \textbf{(iii) repeatable, HCI-relevant diagnostic evaluation}. DailyDroid comprises 75 tasks across 25 widely used applications spanning five realistic categories and three difficulty levels. We evaluate agents under both \textbf{text-only (screentext)} and \textbf{multimodal (screentext + screenshot)} conditions within the same task suite, enabling controlled comparisons of capability, efficiency, and cost. Beyond success rates, we provide step count, execution time, and monetary cost, and we contribute a \textbf{structured failure taxonomy} with standardized failure labels and traces to support reproducible failure characterization and cumulative analysis across future agent iterations. Together, DailyDroid supports not only “which setting performs better,” but also \emph{what human-centered costs that performance entails}, informing the design of future mobile agents, permission models, and mobile UI accessibility practices.

\begin{table}[h!]
\centering
\caption{The DailyDroid Benchmark.}
\Description{Table showing that the DailyDroid benchmark spans 25 apps across five categories, with three tasks per app for a total of 75 tasks.}
\label{tab:benchmark_apps}
\begin{tabular}{@{}lp{6.2cm}cc@{}}
\toprule
\textbf{Task Category} & \textbf{Apps} & \textbf{Tasks per App} & \textbf{Total Tasks} \\
\midrule
Productivity \& Tools     & Calendar, Email, Notes, Files, Tasks          & \multirow{5}{*}{\parbox{2.5cm}{\centering 3 (Simple, Medium, Hard)}} & 15 \\
Utilities \& System Tasks & Contact, Clock, Settings, Weather, Calculator &                                                                     & 15 \\
Information Access      & Wikipedia, Chrome, OsmAnd, Maps, News            &                                                                     & 15 \\
Media \& Entertainment    & Spotify, YouTube, Podcasts, Books, Gallery    &                                                                     & 15 \\
Communication           & Facebook, Reddit, Meeting, Messages, Instagram  &                                                                     & 15 \\
\midrule
\textbf{Total} & \textbf{25 } & \textbf{3} & \textbf{75} \\
\bottomrule
\end{tabular}
\end{table}

\subsubsection{Formation}
As summarized in Table~\ref{tab:benchmark_apps}, DailyDroid contains 75 tasks across 25 apps at three difficulty levels. We designed DailyDroid to be (1) authentic to everyday smartphone interactions, (2) suitable for controlled screentext vs. screenshot comparisons, and (3) compatible with repeatable, diagnostic evaluation from an HCI perspective.

\textbf{Step 1: Deriving task categories from large-scale app usage signals.}
To ground the benchmark in real-world phone use, we first collected category-level download statistics from the Google Play ecosystem (via AppBrain) as of March 2025\footnote{\url{https://www.appbrain.com/stats/android-market-app-categories}}. We then used an LLM as an assistive tool to cluster and summarize common smartphone activity patterns reflected by these categories (e.g., communication, productivity, and media usage). Importantly, the LLM output was treated as a draft coding suggestion, and the final set of categories was determined by author review and consolidation. This process resulted in five broad task categories that are both frequent in daily use and diverse in interaction patterns: \textit{Productivity \& Tools}, \textit{Utilities \& System Tasks}, \textit{Information Access}, \textit{Media \& Entertainment}, and \textit{Communication}.

\textbf{Step 2: Selecting representative apps per category.}
Within each category, we selected five representative apps (25 total) using three criteria: (i) high real-world adoption and broad user base (to avoid niche workflows), (ii) interaction diversity (to cover different UI patterns such as lists, search, forms, media controls, and settings), and (iii) feasibility for repeatable evaluation (tasks can be executed without paid transactions and without requiring private user content). For reproducibility, all apps were installed from APKPure\footnote{\url{https://apkpure.com/}}, and we recorded the package identifiers and app versions used in our experiments (reported in Appendix~\ref{sec:dailydroid}).

\textbf{Step 3: Designing tasks that reflect everyday intents and support diagnosis.}
For each app, we designed three tasks at increasing difficulty (Simple, Medium, Hard), yielding 75 tasks in total (5 categories $\times$ 5 apps $\times$ 3 difficulty levels). Tasks were authored to reflect common user intents (e.g., search, configure, create, organize, share) and to surface realistic sources of friction (e.g., navigation depth, ambiguous UI labels, multi-step confirmation, dynamic content, permission prompts). We also avoided tasks that require sensitive personal data or irreversible actions (e.g., payments), enabling safe and repeatable execution. We pilot-tested tasks to ensure they were executable end-to-end and removed or revised tasks that depended on unstable, time-sensitive, or highly personalized content.

\textbf{Step 4: Defining difficulty levels as an operational rubric.}
We operationalized difficulty based on interaction complexity rather than subjective judgment: \textit{Simple} tasks typically require a short linear sequence within one screen or one feature; \textit{Medium} tasks involve multi-screen navigation and at least one decision point (e.g., choosing among options, correcting an input); \textit{Hard} tasks include longer horizons with multiple decision points, higher ambiguity, or dependencies on system/app states (e.g., toggling settings that change subsequent UI behavior). This rubric is documented alongside each task in Appendix~\ref{sec:dailydroid} to support reproducibility and future extensions.

\textbf{Step 5: Experimental matrix for modality and model comparison.}
Each task was executed under two input modalities, text-only (screentext) and multimodal (screentext + screenshot), to isolate the impact of representation on mobile task automation. To assess how current LLM capability, including recent reasoning improvements, interacts with modality choice, we evaluated two OpenAI models: GPT-4o (baseline multimodal) and o4-mini (reasoning-oriented). Overall, the benchmark yields $75 \times 2 \times 2 = 300$ runs. Appendix~\ref{sec:dailydroid} provides the full benchmark specification and per-task descriptions.

\subsection{Evaluation}

We evaluate DailyDroid with a standardized protocol that separates (i) experimental preparation, (ii) controlled execution with consistent metrics, and (iii) post-hoc validation and diagnostic coding.

\subsubsection{Before Evaluation}
We adopted and extended AutoTask \citep{pan2023autotask}, an open-source Android mobile agent framework\footnote{\url{https://github.com/BowenBryanWang/AutoTask}}. AutoTask implements common agent components (e.g., perception, planning, execution, reflection) and includes backtracking to recover from mistaken actions. To isolate core agent capability, we disabled AutoTask's contextual knowledge injection and memory modules.
We ran all experiments on an Apple M2 Pro machine (16\,GB RAM) using an Android emulator (Pixel XL, Android 11.0) for consistency and reproducibility. We installed AutoTask's action record application, which leverages Android Accessibility Service\footnote{\url{https://developer.android.com/reference/android/accessibilityservice/AccessibilityService}} to capture UI structure and interaction logs, and used Android Debug Bridge (ADB) for emulator control and data collection.

\begin{figure}[h]
  \centering
  \includegraphics[width=\linewidth]{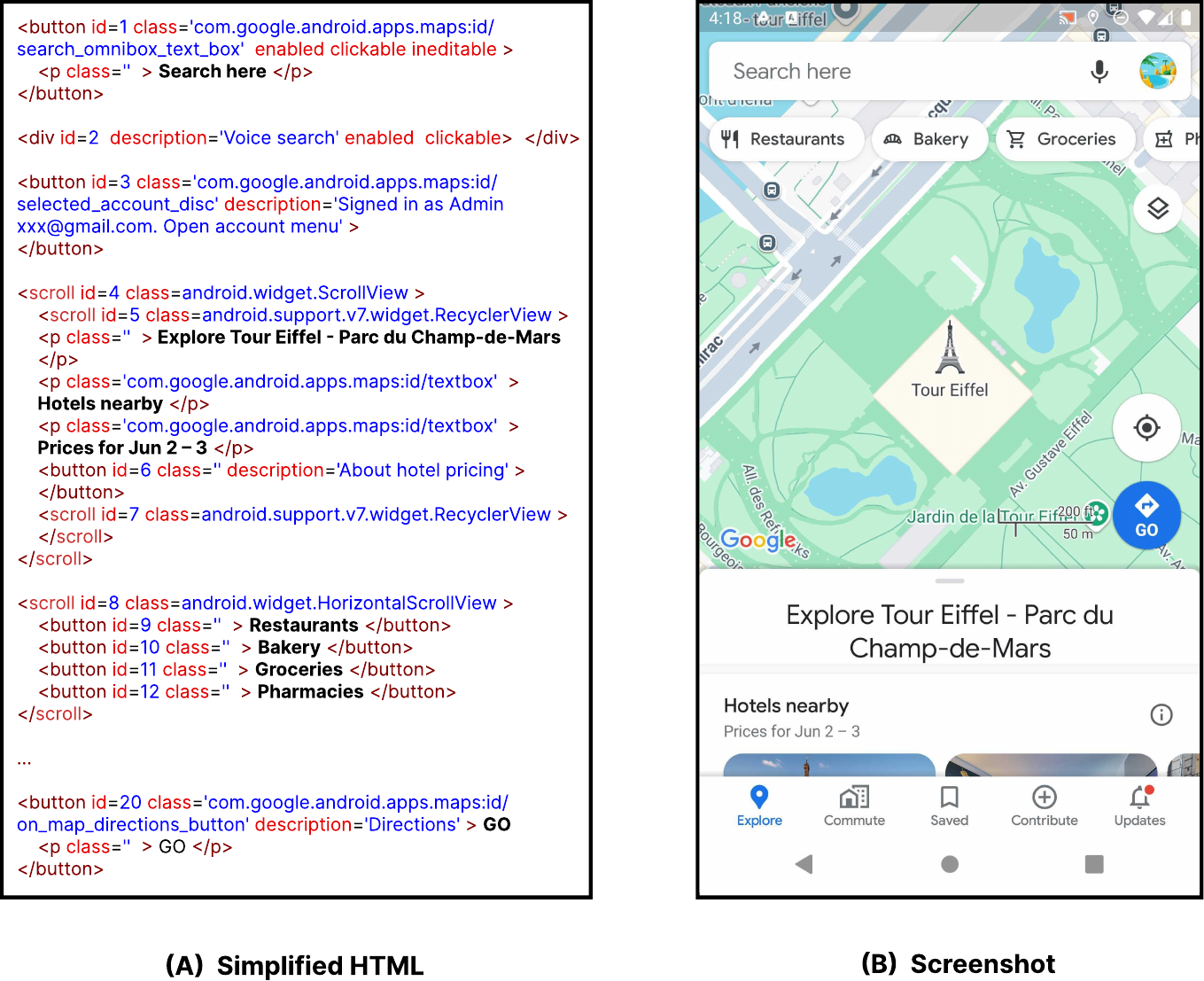}
  \caption{Comparison of screen representations. (A) Simplified HTML showing the structured text representation of the Google Maps interface. (B) Corresponding screenshot of the Google Maps interface displaying the same content visually.}
  \Description{Two panels compare screen representations. (A) Shows simplified HTML code capturing the structure of the Google Maps interface, including elements like search box, landmarks, and nearby categories such as Restaurants, Bakery, Groceries, and Pharmacies. (B) Shows the corresponding screenshot of Google Maps with the same elements displayed visually, including the map centered on the Eiffel Tower and buttons for categories and directions.}
  \label{figure2}
\end{figure}

\subsubsection{During Evaluation}

\textbf{Input modalities.} AutoTask originally supports only a structured screentext representation extracted from the UI tree. We implemented screenshot capture at each step and formed a multimodal condition by pairing screenshots with the same screentext. In our experiments, \textit{text-only} uses screentext only, while \textit{multimodal} uses screentext + screenshot. Figure~\ref{figure2} compares these representations.
AutoTask transforms the extracted UI tree into simplified HTML-like elements with identifiers to preserve hierarchical structure for downstream reasoning and action grounding. All runs were executed with the same emulator configuration and logging pipeline to ensure comparability across modalities and models.

\textbf{Models.} We evaluated two OpenAI models: GPT-4o (baseline) and o4-mini (reasoning-oriented). Combined with two modalities, this yields $75 \times 2 \times 2 = 300$ runs.

\textbf{Run procedure and termination.} Each run proceeds in a loop of (1) observing the current screen representation, (2) generating the next action, (3) executing the action on the emulator, and (4) reflecting on progress. We cap each run at a maximum of 10 agent steps to bound cost and avoid infinite loops.

\textbf{Metrics.} We report a compact but comprehensive set of metrics covering effectiveness, efficiency, and resource use:
\begin{itemize}
    \item \textbf{Task Success (binary)}: whether the agent completes the task as specified.
    \item \textbf{Time}: \textit{Total Time} (end-to-end runtime), \textit{LLM API Time} (time waiting for LLM responses), and \textit{Effective Time} (device-side execution time), where Effective Time $=$ Total Time $-$ LLM API Time.
    \item \textbf{Step Count}: number of executed UI actions (e.g., tap, type, scroll, back) taken by the agent.
    \item \textbf{Token Usage}: total input + output tokens across all LLM calls in a run.
    \item \textbf{Cost (USD)}: token usage converted to monetary cost using OpenAI API pricing as of June 2025 \citep{openai:modelcost}. We apply per-model input/output rates to the logged token counts (GPT-4o: \$2.5/1M input, \$8/1M output; o4-mini: \$0.15/1M input, \$0.6/1M output).
\end{itemize}

\subsubsection{After Evaluation}
After completing all 300 runs, we reviewed logs for each run, including screentext and screenshots, step-level action traces, terminal outputs, and all LLM prompts and responses. We extracted quantitative metrics (time, steps, tokens, cost) and manually verified task outcomes. This audit was necessary because the reflection module can occasionally mark runs as successful due to hallucination or partial completion.

To analyze failures, we employed an inductive coding approach inspired by Grounded Theory \citep{glaser2017discovery}. The first author systematically reviewed all failed runs and iteratively developed failure categories from the data. Initial codes were assigned based on recurring failure patterns observed in logs and model outputs, then refined and consolidated as analysis progressed into higher-level categories (system-level vs.\ agent-level failures). This bottom-up process allowed failure types to emerge from the dataset rather than being imposed by a predefined scheme. The resulting taxonomy is reported in Table~\ref{tab:failure_handbook}, and per-run coding notes were recorded to support repeatable diagnostic analysis.

\section{Results}
\begin{table*}[h]
\centering
\caption{Failure Handbook and a comparative failure analysis of text-only and multimodal modality in our study. Text denotes text-only and multi denotes to multimodal.}
\Description{Table showing a failure handbook of system-level and agent-level error types on the left, with their frequencies across GPT-4o and o4-mini in text-only and multimodal conditions on the right.}
\label{tab:failure_handbook}
\small 
\begin{tabularx}{\textwidth}{@{}m{1.4cm}p{1.8cm}Xrrrr@{}}
\toprule
\multicolumn{3}{c}{\textbf{Failure Handbook}} & \multicolumn{4}{c}{\textbf{Our Study}} \\
\cmidrule(lr){1-3} \cmidrule(lr){4-7}

\multirow{2}{*}{\textbf{Category}} & \multirow{2}{*}{\textbf{Subcategory}} & \multirow{2}{*}{\textbf{Description}} & \multicolumn{2}{c}{\textbf{GPT-4o}} & \multicolumn{2}{c}{\textbf{o4-mini}} \\
\cmidrule(lr){4-5} \cmidrule(lr){6-7}
& & & \textbf{Text (\%)} & \textbf{Multi. (\%)} & \textbf{Text (\%)} & \textbf{Multi. (\%)} \\
\midrule

\multirow{5}{*}{\begin{tabular}[c]{@{}c@{}}System-\\Level\end{tabular}}
& UI Retrieval & Fails to fetch the entire application UI. & \textbf{25} (\textbf{33.3}) & \textbf{25} (\textbf{33.3}) & \textbf{23} (\textbf{30.7}) & \textbf{23} (\textbf{30.7}) \\
& UI Parsing & Retrieves UI but fails to parse or identify the specific components required to proceed. & 3 (4.0) & 3 (4.0) & 4 (5.3) & 5 (6.7) \\
& UI Logic & UI design is ambiguous or counter-intuitive, even for humans. & 1 (1.3) & 1 (1.3) & 0 & 0 \\
& Execution & Identifies correct action but fails to execute. This includes errors in performing operations or lacking action space. & 3 (4.0) & 3 (4.0) & 3 (4.0) & 2 (2.7) \\
& \textbf{Overall} & & \textbf{32 (42.7)} & \textbf{32 (42.7)} & \textbf{30 (40.0)} & \textbf{30 (40.0)} \\
\midrule

\multirow{9}{*}{\begin{tabular}[c]{@{}c@{}}Agent-\\Level\end{tabular}}
& LLM Prediction & Understands the screen but makes wrong predictions. & \textbf{10} (\textbf{13.3}) & \textbf{7} (\textbf{10.8}) & \textbf{7} (\textbf{9.3}) & 3 (4.0) \\
& LLM Reflection & Fails to recognize task completion or step mistakes. & 2 (2.7) & 1 (1.3) & 2 (2.7) & 2 (2.7) \\
& Reaching Max Step & Exceeds step limit due to inefficiency, looping, or a correct but slow path. & 0 & 1 (1.3) & 3 (4.0) & \textbf{7} (\textbf{9.3}) \\
& Impossible Task & Cannot complete task due to intentional task design or ambiguities in the task prompt. & 1 (1.3) & 1 (1.3) & 1 (1.3) & 1 (1.3) \\
\cmidrule(lr){2-7}
& UI Retrieval & \multirow{4}{=}{Similar to the one in System-Level Failure, but it terminates during execution or upon reaching the maximum step, rather than prematurely.} & 2 (2.7) & 2 (2.7) & 2 (2.7) & 2 (2.7) \\
& UI Parsing & & 4 (5.3) & 4 (5.3) & 2 (2.7) & 1 (1.3) \\
& UI Logic & & 3 (4) & 3 (4) & 1 (1.3) & 1 (1.3) \\
& Execution & & 1 (1.3) & 0 & 5 (6.7) & 3 (4.0) \\
& \textbf{Overall} & & \textbf{23 (30.7)} & \textbf{19 (25.3)} & \textbf{23 (30.7)} & \textbf{20 (26.7)} \\

\bottomrule
\end{tabularx}
\end{table*}

We analyzed all experiments by first distinguishing between \textbf{Task Success} and \textbf{Task Failure}. Failed tasks were further examined, classifying them into system-level and agent-level failures. The findings were then consolidated into a \textbf{Failure Handbook} (Table \ref{tab:failure_handbook}). The handbook categorized failures with clear descriptions, providing a practical reference for identifying and diagnosing errors in mobile task automation. Additionally, specific failure distributions were presented alongside the handbook to offer a comprehensive overview of the failure patterns in our study.

System-level failures are critical breakdowns in the underlying infrastructure that invalidate further progress and force early termination. They stem from UI access and execution constraints (UI retrieval, parsing, logic, and action execution) and typically end within 4–5 steps, as human examiners identify fatal blockers and halt the run. Such early termination is critical for saving substantial time and reducing unnecessary experimental overhead. By contrast, agent-level failures arise when the system remains functional but errors occur downstream. They cover a broader set of causes, including the same issues observed at the system level but surfacing later in execution, as well as LLM prediction and reflection errors, reaching the maximum-step limit, and impossible tasks.

\subsection{Task Performance Analysis}
\begin{table*}[htbp]
\centering
\caption{Overall Performance Comparison across Modalities and Models. “Avg.”, "Eff.", and “Tot.” denote average, effective, and total, respectively. Task outcomes are classified as successes or failures, with failures further subdivided into system-level and agent-level categories. Evaluation metrics correspond to the results of task success reported in each row.}
\Description{Table comparing GPT-4o and o4-mini on text-only and multimodal tasks, showing GPT-4o as faster but costlier, and o4-mini as slower but cheaper, with multimodal inputs improving success rates.}
\label{tab:overall_performance_bold}
\resizebox{\textwidth}{!}{
\begin{tabular}{@{}llccccccccc@{}}
\toprule
\textbf{Model} & \textbf{Modality} & \textbf{\shortstack{Task \\ Success}} & \textbf{\shortstack{System-Level \\ Failure}} & \textbf{\shortstack{Agent-Level \\ Failure}} & \textbf{\shortstack{Avg. Total \\ Time (s)}} & \textbf{\shortstack{Avg. Eff. \\ Time (s)}} & \textbf{\shortstack{Avg. API \\ Delay(s)}} & \textbf{\shortstack{Avg. \\ Steps}} & \textbf{\shortstack{Avg. LLM \\ Cost (USD)}} & \textbf{\shortstack{Tot. LLM \\ Cost (USD)}} \\
\midrule
\multirow{3}{*}{GPT-4o} 
& Text-only & 20 (26.7\%) & 32 (42.7\%) & 23 (30.7\%) & \textbf{141.92} & 72.75 & \textbf{69.17} & 6.30 & 0.0947 & 1.8939 \\
& Multimodal & \textbf{24 (32.0\%)} & 32 (42.7\%) & 19 (25.3\%) & 144.86 & \textbf{71.36} & 73.51 & \textbf{5.71} & 2.4593 & 59.0226 \\
\cmidrule(l){2-11}
& Overall & 44 (29.3\%) & 64 (42.7\%) & 42 (28.0\%) & 143.52 & 71.99 & 71.53 & 5.98 & 1.3845 & 60.9165 \\
\midrule
\multirow{3}{*}{o4-mini} 
& Text-only & 22 (29.3\%) & 29 (38.7\%) & 24 (32.0\%) & 204.97 & 85.28 & 119.69 & 7.36 & \textbf{0.0077} & \textbf{0.1702} \\
& Multimodal & \textbf{25 (33.3\%)} & 30 (40.0\%) & 20 (26.7\%) & 220.39 & 86.43 & 133.96 & 7.00 & 0.1906 & 4.7661 \\
\cmidrule(l){2-11}
& Overall & 47 (31.3\%) & 59 (39.3\%) & 44 (29.3\%) & 213.17 & 85.89 & 127.28 & 7.17 & 0.1050 & 4.9362 \\
\bottomrule
\end{tabular}
}
\end{table*}
As detailed in Table \ref{tab:overall_performance_bold}, experiments were compared across modalities and models using multiple metrics. We analyzed performance differences across five dimensions: task completion, modalities, execution time, LLM costs, and task failure.

In terms of task completion, experiments with the o4-mini model slightly outperformed those with GPT-4o. Across the three main metrics, task success, system-level failure, and agent-level failure, results varied by no more than $\pm$3\%, with o4-mini achieving the highest task success rate at 31.3\% compared to 29.3\% for GPT-4o. In addition, o4-mini required approximately 67\% more total time, and took about one more step on average but incurred approximately 13 times lower LLM cost.

In terms of modalities, with o4-mini, experiments with the multimodal outperformed those with the text-only. The task outcome differences were relatively small, with the multimodal setting achieving a 4\% higher task success rate, a 1.3\% higher system-level failure rate, and a 5.7\% lower agent-level failure rate. However, the model API waiting time increased significantly in the multimodal, largely due to the additional tokens introduced by screenshots. With the help of screenshots, the multimodal required 0.36 fewer steps on average, suggesting reduced execution errors. This performance gain, came at a considerable cost, resulting in approximately 25 times higher LLM costs. In addition, a similar trend was observed with GPT-4o, where the multimodal achieved a 5.7\% higher task success rate, reduced the number of steps by approximately 0.6, but incurred about 26 times higher cost. 

In terms of execution time, the multimodal modality generally took longer than the text-only modality due to the additional overhead of uploading images. Across models, time costs increased notably for o4-mini, driven by longer reasoning processes. Interestingly, total running times were still longer than expected, with the fastest configuration-text-only with GPT-4o-averaging 22.5 seconds per step. When excluding API latency, the effective time for this configuration was 11.5 seconds per step. Compared to other benchmarks, one intuitive answer is the challenge of our in-the-wild benchmark. After further looking closer to experimental logs, another three main contributed to this discovery. First, the emulator's front-end required delays to synchronize with the back-end and load subsequent pages, as the perception module in AutoTask frequently fetched the UI. Second, the orchestration of multiple modules within the mobile agents introduced inter-module communication overhead. Finally, the experiments ran on an Apple M2 Pro with 16 GB memory, where communication between the computer and emulator also incurred latency. In terms of LLM costs, GPT-4o with multimodal input incurred the highest expense, with a total cost about 12× higher than o4-mini. This aligns with a common deployment reality: newer model families can be cheaper per token even when they offer stronger capabilities. In our results, per-run averages muted the difference, but aggregating across all runs amplified it, underscoring the value of reporting total cost alongside average cost.

In terms of task failure, overall, system-level failures were the primary source of task failures, accounting for 42.7\%, 42.7\%, 40.0\%, and 40.0\% for GPT-4o (text), GPT-4o (multimodal), o4-mini (text), and o4-mini (multimodal), respectively. Specifically, the most common failure was UI retrieval with over 30\%, though other failures such as UI parsing errors, UI logic, and execution were also observed. In contrast, LLM prediction errors were a major cause of failure in agent-level. With the advanced reasoning model, the prediction error rate significantly decreased—from 13.3\% to 9.3\% in the text-only modality, and from 10.8\% to 4.0\% in the multimodal modality. This indicates that more advanced models, especially when supported by richer visual information, substantially reduce the likelihood of LLM-level mistakes. Another notable finding involves failures due to reaching the maximum step limit. Interestingly, more advanced models were more prone to exceeding the predefined step limit, with an increase of 12\%. We further analyzed the underlying causes of this phenomenon in \ref{maximumstepconstraints}.

\subsection{System-level Failures} For system-level failures, most issues stemmed from LLMs’ difficulty in interpreting certain UI designs, either the entire UI retrieval or the specific components. We observed that such issues occurred primarily in two types of tasks: (1) complexly designed social or media applications that lacked accessible UI elements for LLMs to interact with, (2) operations involving numeric input, such as setting time or performing calculations.

The most common issue stemmed from inaccessible or unavailable UI elements. When creating a Reddit post or viewing an Instagram post, missing UI information forced the agent to explore blindly; even with screenshots providing more visual clues, the task still failed without actionable UI access. Similar results were also found in Wikipedia, Podcasts, Spotify, and YouTube. On the other hand, lacking specific UI components, such as the title box in Calendar, the search bar in Google Books, or the time selector in Clock, posed challenges for the LLMs. For example, in Google Books, the hidden search bar prevented the model from executing the search operation.

Another issue stemmed from unintuitive UI Logic. In highly graphical or numerically oriented apps, LLMs could recognize the interface but struggled with precise actions. For example, when setting a 25-minute timer, the model consistently misinterpreted the numeric input order, predicting `5' before `2' instead of entering them correctly. Such unintuitive designs which often confuses even for human users made tasks especially difficult, since LLMs selected actions from predefined options based on the scoring system rather than provided with prior knowledge. Similar actions can also be found under agent-level failures or on calculation tasks in Calculator discussed in \ref{emergingllm}. Addressing this challenge requires either equipping LLMs with more task-specific usage knowledge or redesigning UIs to be more intuitive for both humans and intelligent agents.

Regarding these system-level failures, we recommend that application developers ensure the UI accessibility and UI designers provide systematic usage guidance within the application (e.g., clear manuals or instructions). This is particularly important for applications designed primarily for human interaction, where strict restrictions on UI access otherwise make mobile automation tasks difficult to implement.

\subsection{Agent-level Failures}
Execution issues appeared in both system-level and agent-level failures. When tasks required specific actions, such as attaching documents or long-pressing UI components to access metadata, the absence of such operations in the predefined action space often triggered failures. Expanding the action space could help mitigate this problem. In some cases, the execution module also deviated from the prediction module’s predicted actions, underscoring the need to strengthen inter-agent communication.

Beyond these failures, LLM-related issues, prediction and reflection, were prominent at the agent level. Although the necessary UI components were available, LLMs occasionally misinterpreted task-relevant elements, leading to incorrect predictions that impeded completion. For example, they confused last name and phone number fields when creating new contacts, or misread location-based instructions in Google Maps, leading to incorrect route searches. With respect to reflection, the framework’s iterative back-and-forth process occasionally left the LLM uncertain about whether a task was complete, especially in multi-step or cross-app scenarios. For instance, when asked to send photos via email or to execute a sequence of calling, hanging up, and then messaging, the model often failed to press the final send button. These errors reveal difficulty in maintaining task progression, with termination frequently triggered by reaching the step limit.

Other agent-level failures include reaching the maximum-step limit and impossible tasks. The former often reflects lengthy or error-prone task progressions (an interesting trade-off between reasoning and step constraints was further discussed in \ref{maximumstepconstraints}), while the latter was rare and typically arose from tasks beyond the agents’ capabilities.

Based on both statistics shown in Table \ref{tab:failure_handbook} and human review of the logical performance comparing LLMs, o4-mini outperformed GPT-4o. Exploring the reasons for agent-level failures, we found GPT-4o mainly failed from direct wrong predictions and misunderstandings, while o4-mini showed more self-correction behavior but still failed due to exceeding step limits or UI issues. Moreover, our analysis revealed a clear distinction in the strengths of the two models. o4-mini demonstrated stronger performance in utility and system-level operations (52\% success rate in Utilities/System and Productivity/Tools tasks), as well as in navigation-oriented activities such as managing Wi-Fi connections, clearing caches, or planning routes in Maps. These tasks often required precise, sequential actions across device settings, where o4-mini shows greater consistency. In contrast, GPT-4o excelled in socially and content-driven contexts (60\% success rate in Information Access, Social/Communication, and Media/Entertainment tasks). It performed better at interacting with social media platforms, handling communication workflows (e.g., messages, calls, emoji use), and engaging with content-rich apps such as Google Books and Facebook. This suggested that o4-mini is better suited for device management and automation, while GPT-4o excels in user-centric, social, and content-rich applications.

\section{Discussion}
\subsection{UI Accessibility}
Our analysis reveals that issues related to UI perception are the primary obstacle in mobile task automation. These errors, originating in the Perception Module, are fundamental; if an agent cannot accurately perceive the UI, any subsequent reasoning by the LLM becomes irrelevant or lucky at best. This is a task that a human would also fail. A closer inspection of the logs for these UI-related failures shows that the root cause often lies in missing or incomplete text content within the screen representation. This gap between the visual reality and the parsed data was particularly evident in our multimodal experiments. On several occasions, the screenshot provided enough visual context for the LLM to correctly identify the next logical step, yet the task was doomed to fail because the target UI elements were absent. In these instances, the agents were aware of what to do but ultimately failed to do it, often continuing to explore fruitlessly until reaching the maximum step limit. Similarly, prior research on UI accessibility for individuals with visual impairments has also highlighted the absence of essential UI components at the app level \citep{al2021systematic, mohit2021Blind}. Examples include missing button labels \citep{ross2018examining} and the lack of alternative text for images in social media applications \citep{morris2016most}. These findings underscore UI accessibility as a persistent challenge in this domain and indirectly reflect the inherent difficulties faced in mobile task automation. 

In general, our findings point to two major obstacles in advancing this field: the complexity of mobile user interfaces and the limitations of current parsing methods. Complex UIs vary substantially in content and structure across applications, devices, and vendors. As highlighted in the W3C mobile accessibility guidelines \citep{mohit2021Blind, w3c_mobile_2015}, providing clear labels for interactive elements is essential. Addressing this critical issue requires multi-level interventions, such as encouraging app developers to thoroughly review UI and accessibility features prior to release, developing general-purpose tools to support app-level accessibility \citep{rodrigues2017context}, and implementing stricter UI accessibility policies by mobile device manufacturers to reduce inconsistencies. A second obstacle lies in parsing methods. The AutoTask framework in our experiment uses a rule-based mapping to transform Android components into an HTML-like structure. While effective on the curated datasets where it was originally validated, such as PixelHelp \citep{li2020mapping} and UGIF \citep{venkatesh2022ugif}, its generalizability to ``in-the-wild'' applications is not guaranteed. This challenge is well-recognized. Seminal research on applying LLMs to mobile UI interaction \citep{wang2023enabling} developed a custom depth-first search algorithm with heuristic rules\footnote{https://github.com/google-research/google-research/tree/master/llm4mobile} to generate screen representation, yet also noted the complexity and unreliability of UI extraction and validated their approach on specific datasets. In addition, the common UI parser relies solely on converting view hierarchy information into an HTML representation, leaving other modalities such as pixels unused. This limitation highlights the unresolved challenge of building a truly universal UI parser.

\subsection{Input Modalities}
Current LLM-based mobile agents largely rely on two fundamental input modalities to understand the screen: textual UI-trees \citep{wang2023enabling, wen2024autodroid} and visual screenshots \citep{Zhang2023AppAgentMA, cheng2024seeclick, baechler2024screenai}. Although researchers recognize their inherent differences, a common trend has been to create a hybrid that fuses both modalities, often in pursuit of higher task success rates \citep{Zhang2023AppAgentMA, rawles2024androidworld}. 

\begin{figure}[h]
  \centering
  \includegraphics[width=\linewidth]{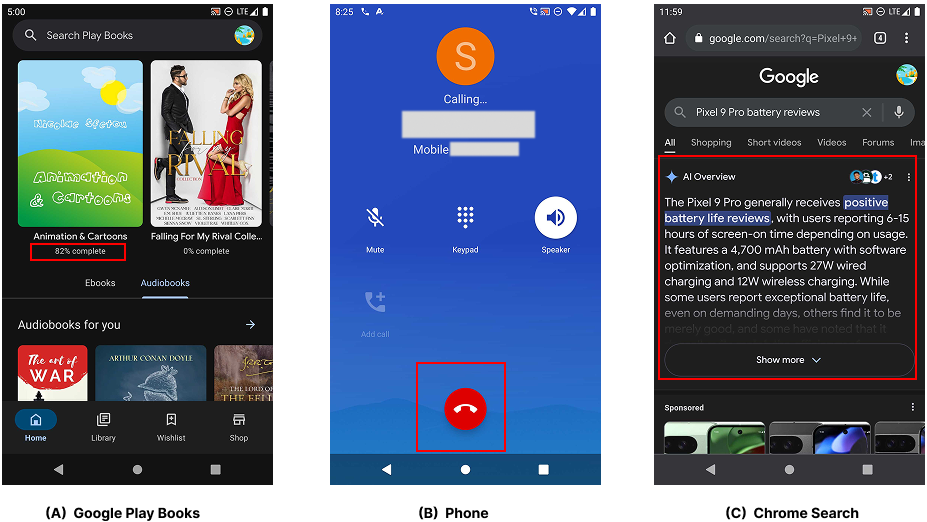}
  \caption{Failed Cases of the Text-only modality. (A) In Google Play Books, the red rectangle highlights the reading progress. (B) In Phone app, the red rectangle highlights the hang-up button. (C) In Chrome search, the red rectangle highlights the AI Overview of the search content.}
  \Description{Three smartphone screenshots showing failed cases for the text-only modality. (A) Google Play Books with a red rectangle highlighting the reading progress bar. (B) Phone app with a red rectangle highlighting the red hang-up button during a call. (C) Chrome Search with a red rectangle highlighting the AI Overview snippet in search results.}
  \label{failedcases}
\end{figure}

To our knowledge, a systematic comparative analysis of these modalities is currently lacking in the literature. As mentioned in Section \ref{screencontent}, each representation has distinct advantages and disadvantages. Our results reveal that the multimodal stems from capturing crucial visual context that is absent from the textual-only. Take some typical examples in Figure \ref{failedcases}:
\begin{itemize}
    \item Simple task: "Open the Google Play Books app and continue reading the last opened book from library." With a screenshot, the model can see the exact reading progress in Google Play Book, whereas the text-only lacks this cue.
    \item Medium task: "Make a phone call to \{people\} and then hang up. Send an emoji with love to this person.". The text-only modality fails to locate the red hang-up button, while the multimodal modality effectively understands the semantics behind this obvious red hang-up icon and proceeds to click the icon. Similarly, when sending an emoji, due to lack of the context of chatting interface in Message app, the text-only one still fails to click the send button though the correct emoji is typed into the chat input field.
    \item Hard task: "Open Chrome and search for reviews of the Pixel 9 Pro battery. Summarize the findings about battery life in 1-2 sentences.". A very difficult or impossible task is designed to evaluate the agent's capability and the difference between two modalities. In the multimodal condition, the model realizes the "AI Overview" function provided by Chrome with the help of screenshot, and then directly take part of the content as the final summarization. However, the text-only captures only the title of "AI Overview", making the model continue to browsing endlessly.
\end{itemize}

Therefore, the multimodal modality makes full use of visual clues and stands out under the following conditions: (1) When tasks are complex, multi-step, or abstract. (2) When task progression depends on non-textual visual cues (icons, colors, and spatial layout), such as weather icons, map pins, and communication tools. (3) When the screentext is incomplete or ambiguous, fusing it with screenshots dramatically boosts task completion rates.

In addition, modality-specific permissions deserve careful consideration, as multimodal inputs yielded marginally higher task success rates. Screentext data require Android accessibility access, whereas screenshot data require screen recording access. Although both modalities capture the same underlying content, their representation differ substantially. Compared with screentext, screenshots are more invasive and sensitive \cite{reeves2020time, teng2024tool, zhang2024autojournaling}. Under privacy concerns, a key design implication is screentext access should be prioritized over multimodal access. This approach reduces permission requests and mitigates privacy risks while still maintaining comparable performance in mobile task automation. On iOS, however, such access is more restricted and typically limited to testing contexts.

\subsection{LLM capabilities}
\subsubsection{Emerging LLM Capabilities}
\label{emergingllm}
Our evaluation reveals that enhanced reasoning abilities in newer LLMs lead to emergent behaviors not observed in baseline models. These include handling more complex logic, adapting to missing UI elements, and flexibly choosing between alternative solution paths. For instance, when direct actions were unavailable, the reasoning model could identify and execute indirect strategies (e.g., leveraging system settings or switching input modes). Such adaptability demonstrates that advanced LLMs are capable of compensating for both limited action spaces and imperfect UI representations, ultimately broadening the range of tasks they can complete successfully.

\subsubsection{Balancing Reasoning and Step Constraints}
\label{maximumstepconstraints}
Although o4-mini showed improved task success over GPT-4o, our analysis of 44 agent-level failures revealed that 10 tasks in o4-mini reached the maximum step limit, a failure mode not observed in GPT-4o. This indicates that o4-mini’s stronger reasoning capabilities lead to frequent self-corrections through backtracking when execution or prediction errors occur, thereby maintaining task coherence. In contrast, GPT-4o tends to persist with errors, offering fewer or no corrections. As a result, despite some task failures, the reasoning-enhanced model achieves a higher overall task completion rate. Extending the step limit could further improve o4-mini’s success, as these tasks show greater potential for recovery compared to GPT-4o.

\subsection{Framework Design}
\subsubsection{Prompt Design} 
In our experimental logs, the reflection module evaluated whether the current step was executed correctly and predicted the next action. Since both the current execution and historical data were provided as context, its predictions were generally accurate. However, we observed frequent mismatches between the reflection module and the prediction module in the following step, particularly in failure cases. Notably, the reflection module often predicted the next action correctly, while the prediction module failed. Upon inspecting the prompt design, we found that the prediction module lacked sufficient context, receiving only the current UI and previous actions as input. This highlights the need for future agent designs to incorporate both current and historical context for more consistent and reliable task execution.

\subsubsection{Agent Action Space} 
The agent’s current action space is restricted to basic operations (click, edit, scroll). However, many tasks in real-world apps require richer actions \citep{xu2024androidlab}, such as long-press, swipe, back, and home that are unavailable. While agents sometimes improvise alternative paths, these workarounds are inefficient and unreliable. This highlights the importance of expanding the action space to cover a broader and more representative set of interactions for robust task automation.

\subsubsection{Step Constraints}
When an execution error occurred, the reflection module triggered the Backtracking module. We observed that for complex tasks, especially when using the reasoning model o4-mini, allocating more time for deliberation improved both task accuracy and self-correction. However, each correction triggered the Backtracking module, consuming additional execution steps early in the process. This behavior was evident in the ten o4-mini failure cases where the model reached the maximum step limit, an issue almost never observed with GPT-4o. Accordingly, the deployment of reasoning models in complex tasks requires revisiting the maximum step threshold to ensure their iterative correction capabilities can be fully leveraged without prematurely reaching task termination.

\subsection{Better App Design}
From the experiments, we noted that a better model does not improve the task performance significantly. It was the system level errors that occupies both over 50\% in the experiments across two models. Therefore, this informs us to think about how to make better mobile applications. Android has a mature ecosystem about UI design\footnote{https://developer.android.com/design/ui/mobile} for developers, ranging from foundations, layout\&content, to behaviors\& patterns. Among them, Material Design is Google's design system that provides guidelines and principles for creating consistent, visually appealing, and user-friendly interfaces across a wide range of devices \citep{Android2024Material, Pandya2024Material}. It covers various aspects of UI design, including layout, typography, color schemes. However, in terms of UI code design guidelines for app automation, Android does not provide it.

Based on our findings, we provide some insights to help improve app design: (1) Declare and generate stable locators in source code. Using unique and descriptive identifiers "android:id" attributes for all interactable UI elements, would be easier for automation tools to accurately target relevant UI components. (2) Operating system needs to be improved for making easier and consistent parsing protocol across different devices. (3) Making apps more accessible for LLMs. Implementing schema markup and structured data for mobile apps, which provide structural data format that LLMs can easily parse and understand. In addition, new standardized APIs for LLM interaction with mobile apps need to be designed, like Model Context Protocol (MCP)\footnote{https://modelcontextprotocol.io/introduction} for standardized contextual information communication between smartphone data and LLMs.

\subsection{Browser-based GUI Agent}
Although desktop browser-based agents—such as Operator \citep{openai:computeragent} from OpenAI, Computer Use \citep{anthropic_claude_2024} from Claude, and Manus \citep{manus2025}—aim to automate repetitive tasks to save time and reduce errors, they operate in fundamentally different environments compared to mobile agents. Desktop browsers typically provide more stable environments, fewer device variations, and faster network connections. As a result, browser-based agents face lower complexity but are limited to interactions within web browsers and web applications. In contrast, mobile agents must handle greater complexity due to a wider range of device features, apps, gesture-based interactions, and operating system restrictions. Essentially, mobile phones contain a browser as well as other apps.

Browser-based agents generally rely on screenshots as input, enabling them to interact with GUI using pre-trained LLMs, while avoiding direct interaction with complex front-end structures like HTML and website APIs \citep{anthropic_claude_2024, openai:computeragent}. Similarly, in the more complex domain of universal computer use, the OSWORLD benchmark provides both screenshots and additional information—such as accessibility trees—as input \citep{xie2024osworld}. Given the capabilities of modern devices, combining both visual (screenshots) and textual inputs is often more effective.

Recent research on mobile pre-trained LLMs \citep{li2025ferret, liu2024mobilellm} and on-device LLMs \citep{xu2024on} further highlights the potential for mobile task automation. Despite ongoing progress, there remains significant room for improvement in task automation on mobile platforms.

\section{Limitations and Future Work}
In this section, we discuss our study's limitations. In this work, solely screenshot-based modality was not evaluated, as AutoTask framework did not support. This limitation presents a critical challenge in grounding the LLM's perception into actionable commands. To address this, frameworks like AppAgent \citep{Zhang2023AppAgentMA} annotate interactable UI elements with unique identifiers, which allows the LLMs to ground its visual analysis by predicting an action on a specific identifier. However, we did not develop a similar function integrated in AutoTask, since it is beyond the scope of this work. Future work could further investigate this modality, yielding more comprehensive and robust insights into its impact on mobile task automation.

Furthermore, the UI function availability is witnessed unstable. Before finalizing the specific emulator, emulator with latest Android API and Pixel 8 Pro was explored. However, Android accessibility service cannot fetch the detailed UI trees at most system-level apps (Appendix \ref{sec:wrongUI}). Therefore, we had to roll back to the current emulator with elder Android API mentioned in AutoTask. An additional issue appears that the Google Play store was not compatible with the older version emulator, largely raising testing difficulty. Future research should addresses the UI accessibility issues to minimize biases introduced by emulators, devices and applications, thereby enabling more consistent reproducibility.

Additionally, to ensure robustness and mitigate random fluctuations, each task would be ideally executed multiple times. However, because our benchmark comprised a broad set of in-the-wild Android applications rather than a small, controlled set, a multi-run protocol was computationally prohibitive due to the substantial overhead in testing and analysis. We contend that the breadth of our evaluation across numerous diverse tasks sufficient exposed the systemic failure modes and common challenges of current frameworks and models, even without repeated trials. Further study could standardize benchmarks into a more comprehensive and controlled form, drawing on examples PixelHelp \citep{li2020mapping} and UGIF \citep{venkatesh2022ugif} in this field.

\section{Conclusion}
In this work we proposed and evaluated the DailyDroid benchmark for assessing mobile agents in mobile task automation, focusing on input modalities across different LLMs. Our experiments identified the causes and contexts of mobile agent failures, consolidating these findings into a failure handbook for reference. We further examined an discussed issues related to UI accessibility, input modalities, LLM capabilities, and framework or agent design. In addition, we explored principles for designing better applications and compared our findings with similar browser-based GUI agents, offering design implications for future mobile agents, smartphone applications, and UI development. Collectively, these insights advance the understanding of common and critical challenges in mobile task automation, paving the way toward more intelligent personal assistants on smartphones.

\bibliographystyle{ACM-Reference-Format}
\bibliography{Ref}

\clearpage 
\appendix
\section{Appendix A: The DailyDroid Benchmark} \label{sec:dailydroid}

\begin{longtable}{
@{}
>{\raggedright\arraybackslash}p{0.2\linewidth} 
>{\raggedright\arraybackslash}p{0.15\linewidth}
>{\raggedright\arraybackslash}p{0.1\linewidth}
>{\raggedright\arraybackslash}p{0.55\linewidth} 
@{}
}
\caption{The DailyDroid Benchmark} \label{tab:benchmark_grouped} \\
\toprule
\shortstack[l]{\textbf{Task (Sub) Category}} & \shortstack[l]{\textbf{App}} & \shortstack[l]{\textbf{Task Level}} & \shortstack[l]{\textbf{Task Prompt}} \\
\midrule
\endhead

\textbf{Productivity \& Tools} & & & \\

\multirow{3}{=}{\shortstack[l]{Managing Calendar}}& \multirow{3}{*}{Google Calendar} & simple & "Add an event `Doctor Appointment' tomorrow at 2 PM."\\
& & medium & "Schedule a `Team Lunch' for next Friday from 12:30 PM to 1:30 PM at `Cafe Central'."\\
& & hard & "Create a recurring event titled `Project Sync' every Tuesday at 9:00 AM for 45 minutes. Add participant `\{people\}@gmail.com' and set a reminder 15 minutes before." \\
\midrule 

\multirow{3}{=}{\shortstack[l]{Email Tasks}}
& \multirow{3}{*}{Gmail} & simple & "Compose an email to `friend@example.com' with subject `Lunch?' and body `Want to grab lunch tomorrow?'"\\
& & medium & "Open the oldest Email from a keyword search, `GitHub'."\\
& & hard & "Compose a new email to `\{people\}@gmail.com'. Subject: Project Updates. Body: Hi \{people\}, Please find the weekly updates. Remember to cc `\{people\}@gmail.com'." \\
\midrule

\multirow{3}{=}{\shortstack[l]{Note Taking}}
& \multirow{3}{*}{Google Keep Notes} & simple & "Create a new note titled `Meeting Notes' with the text `Discussed project timeline'."\\
& & medium & "Create a new note titled `Meeting Ideas'. Add the following bullet points: `- Discuss Q3 goals', `- Brainstorm new features', `- Assign action items'. Then, pin this note."\\
& & hard & "Create a note titled `Travel Plans'. Add a checklist... Attach the most recent image from the gallery to the note. Add the tag `\#travel'." \\
\midrule

\multirow{3}{=}{\shortstack[l]{File Management}}
& \multirow{3}{*}{Files} & simple & "Open the file manager and navigate to the `Documents' folder."\\
& & medium & "Find a file named `report\_final.pdf' in the `Documents' folder and rename it to `report\_final\_v2.pdf'."\\
& & hard & "Search for all files within this week, sort them by size in descending order, and delete the largest file if it exceeds 10MB." \\
\midrule

\multirow{3}{=}{\shortstack[l]{Managing TODOs}}
& \multirow{3}{*}{Google Tasks} & simple & "Add a new task with a title `Call \{people\}' and save it."\\
& & medium & "Add a new task with a title `Finish report', due tomorrow with a reminder 'later today'."\\
& & hard & "Sort the tasks by day, mark all the tasks before today as done, and unmark the completed task due by tomorrow."\\
\midrule

\textbf{Utilities \& System Tasks} & & & \\

\multirow{3}{=}{\shortstack[l]{Create Contact}}
& \multirow{3}{*}{Contact} & simple & "Create a new contact, `\{people\}', `\{phone number\}', `\{people\}@gmail.com'"\\
& & medium & "Create a new contact: First Name `\{people\}', Mobile Phone \{phone number\}, Work Email `\{people\}@business.com', Company `Tech Solutions Inc.'."\\
& & hard & "Create a new contact: First Name `\{people\}', Last Name `\{people\}', Mobile Phone `\{phone number\}', Work Email `\{people\}@gmail.com', add a Note `Met at the conference'. Add a profile picture by selecting an existing image from the gallery."\\
\midrule

\multirow{3}{=}{\shortstack[l]{Setting Time}}
& \multirow{3}{*}{Clock} & simple & "Set an alarm for 6:30 AM"\\
& & medium & "Set a timer for 25 minutes and label it `Pomodoro'."\\
& & hard & "Create a new alarm for 7:15 AM this Sunday. Set the alarm sound to `BeeBeep' and enable vibration. Name the alarm 'Work Wake Up'."\\
\midrule

\multirow{3}{=}{\shortstack[l]{Adjust System Settings}}
& \multirow{3}{*}{Settings} & simple & "Disable Bluetooth."\\
& & medium & "Adjust the screen brightness to 50\% and turn off dark theme automatically."\\
& & hard & "Navigate to Wi-Fi settings. Disconnect the network named 'AndroidWifi', and forget this network."\\
\midrule

\multirow{3}{=}{\shortstack[l]{App Management}}
& \multirow{3}{*}{Weather} & simple & "Pause the app `Weather' for the rest of the day."\\
&  & medium & "Open the File, search for the apk name `Weather', and install it."\\
&  & hard & "Go to the list of installed apps in settings. Find the `Weather' app and clear its cache."\\
\midrule

\multirow{3}{=}{\shortstack[l]{Calculation}}
& \multirow{3}{*}{Calculator} & simple & "Open the Calculator app and perform a basic calculation, $3 \times 5$."\\
& & medium & "Open the Calculator app and perform a two-step calculation  $50 + 25 \times 3$."\\
& & hard & "Switch the Calculator app to scientific mode, then compute a more complex expression using scientific functions, sin(30) + $5^2$."\\
\midrule

\textbf{Information Access}  & & & \\

\multirow{3}{=}{\shortstack[l]{Information Retrieval}}
& \multirow{3}{*}{Wikipedia} & simple & "Open the Wikipedia app and navigate to `Claude Shannon', then read the Quick Facts about him."\\
& & medium & "Open the Wikipedia app, search for `Michael Bloomberg', tap on the link to `Mayor of New York City', and read the 'History' section."\\
& & hard & "Search `OpenAI' in Wikipedia app, jump to `Initial motivation' through the table of contents, and then read this section."\\
\midrule

\multirow{3}{=}{\shortstack[l]{Google search}}
& \multirow{3}{*}{Chrome} & simple & "Google Search the web for `capital of Australia'."\\
& & medium & "Open browser, Find the current weather forecast for Melbourne for the next 3 days."\\
& & hard & "Open Chrome and search for reviews of the Pixel 9 Pro battery. Summarize the findings about battery life in 1-2 sentences."\\
\midrule

\multirow{3}{=}{\shortstack[l]{Navigation}}
& \multirow{3}{*}{OsmAnd} & simple & "Find the location of Eiffel Tower in maps."\\
& & medium & "Search for nearby attractions to `Eiffel Tower', and click the first one."\\
& & hard & "Plan a walking between `Eiffel Tower' and `Louvre Museum' avoiding ferries."\\
\midrule

\multirow{3}{=}{\shortstack[l]{Find Business Info}}
& \multirow{3}{*}{Google Maps} & simple & "Find the phone number for `Domino Pizza' nearby."\\
& & medium & "Find restaurants near me that are open now, serve Italian food, and have a rating of 4 stars or higher."\\
& & hard & "Find the opening hours for the `National Gallery of Victoria' in Melbourne, Australia for this coming Saturday. Check if they have wheelchair access mentioned on their details page or website."\\
\midrule

\multirow{3}{=}{\shortstack[l]{Reading News}}
& \multirow{3}{*}{Google News} & simple & "Open Google News and click the first news."\\
& & medium & "Within the news app, use the search to find articles about ChatGPT, select the second article from the results."\\
& & hard & "In the news app, select an interesting article from `World' and save it for later reading."\\
\midrule

\textbf{Media \& Entertainment}  & & & \\

\multirow{3}{=}{\shortstack[l]{Playing Music}}
& \multirow{3}{*}{Spotify} & simple & "Play the song `Blinding Lights'."\\
& & medium & "Play the album `Folklore' by Taylor Swift on shuffle."\\
& & hard & "Create a new playlist named `Everyday Chill Vibes', add the songs `Weightless' by Marconi Union and `Teardrop' by Massive Attack, then play the song `APT. - Piano Version' by Cozy Rabbit."\\
\midrule

\multirow{3}{=}{\shortstack[l]{Watching Video}}
& \multirow{3}{*}{YouTube} & simple & "Open the YouTube, type `Perfect' by Ed Sheeran into the search bar, select and play it."\\
& & medium & "While watching the `Perfect' by Ed Sheeran, tap the like button and subscribe to the channel."\\
& & hard & "After playing the `Perfect' by Ed Sheeran, scroll down and post a comment on the video, 'Amazing! Thanks for sharing!', then add the video to your Watch Later playlist."\\
\midrule

\multirow{3}{=}{\shortstack[l]{Listening Podcast}}
& \multirow{3}{*}{Google Podcasts} & simple & "Search for the podcast 'This American Life'."\\
& & medium & "Find the podcast `Radiolab' and play the latest episode."\\
& & hard & "Subscribe to the podcast `Stuff You Should Know'. Download the episode titled `How LEGOs Work' for offline listening. Add the episode `The Case of the Missing Hit' to the queue."\\
\midrule

\multirow{3}{=}{\shortstack[l]{Reading Book}}
& \multirow{3}{*}{Google Play Books} & simple & "Open the Google Play Books app and continue reading the last opened book from library."\\
& & medium & "Open Google Play Books, search the `Catch me If you can', and read the first page."\\
& & hard & "Open Google Play Books, search for books by `Brandon Sanderson'. Filter the results to show only `Ebooks'. Add the first book result to your wishlist. Then, go back to your library, open 'The Hobbit', navigate to Chapter 1 (An Unexpected Party)."\\
\midrule

\multirow{3}{=}{\shortstack[l]{Managing Photos}}
& \multirow{3}{*}{Gallery} & simple & "Open the gallery app and view the most recent photo."\\
& & medium & "Find all photos taken four years ago and add them to a new album called `Vintage memory'."\\
& & hard & "Select the third photo in the Gallery. Apply an `auto-enhance' filter if available, then share the enhanced photo via email to `\{people\}@gmail.com' with the subject `Beautiful pic in South Africa'."\\
\midrule

\textbf{Communication}  & & & \\

\multirow{3}{=}{\shortstack[l]{Sharing Story}} & \multirow{3}{*}{Facebook} & simple & "Open the Facebook app, create a new post, `SayHi from my first facebook' and publish it."\\
& & medium & "Compose a post on facebook: tap the compose button, type `Sayhi' with a hashtag,`\#morning', attach a scenery image from the gallery, then post it."\\
& & hard & "On Facebook, check your latest post, like it and leave a comment, `what a beautiful place', and share with your Feed and friend with `My recent travel'."\\
\midrule

\multirow{3}{=}{\shortstack[l]{Social Forum}}
& \multirow{3}{*}{Reddit} & simple & "Open Reddit app, navigate to RayBanStories, and open the lastest post."\\
& & medium & "Open Reddit app, find the lastest post and add a comment `Sounds good!' in RayBanStories."\\
& & hard & "Open Reddit app, create a new post on RayBanStories, enter the title `MetaAI Feedback' and 'What do you think of MetaAI?', then submit the post."\\
\midrule

\multirow{3}{=}{\shortstack[l]{Online Meeting}}
& \multirow{3}{*}{Google Meeting} & simple & "Open Google Meet and start a new instant meeting."\\
& & medium & "Schedule a new meeting in Google Meet titled `Team Sync' for tomorrow, April 14th, 2025, at 11:00 AM."\\
& & hard & "Schedule a new meeting in Google Meet titled 'Client Presentation' for next Friday at 2:00 PM. Invite `\{people\}@example.com' and `\{people\}.lead@example.com'. Ensure the meeting setting allows anyone with the link to join."\\
\midrule

\multirow{3}{=}{\shortstack[l]{Contact}}
& \multirow{3}{*}{Messages} & simple & "Send a message to \{people\} to ask `where should we meet at dinner time?'"\\
& & medium & "Make a phone call to \{people\} and then hang up. Send a emoji with smile."\\
& & hard & "Share your current location to \{people\}. Then, send a message, `This is the lecture slide' and attach the lecture pdf document in File."\\
\midrule

\multirow{3}{=}{\shortstack[l]{Social Media}}
& \multirow{3}{*}{Instagram} & simple & "Open Instagram, scroll your feed and double-tap to like the latest post."\\
& & medium & "Create a new post on Instagram: tap the create button, select an existing photo from the gallery, add a smile emoji, and then share to the feed."\\
& & hard & "Share a post on Instagram with a friend: tap the create button, select a photo, apply an auto-filter, add a caption `Share my life with my friend \{people\}' with hashtag `\#chill', tag `@\{people\}' in the post, and then share to the feed."\\

\bottomrule
\end{longtable}

\clearpage
\section{Appendix B: Example of Incorrect UI Extraction in Emulator} 
\label{sec:wrongUI}
\begin{figure}[h]
  \centering
  \includegraphics[width=\linewidth]{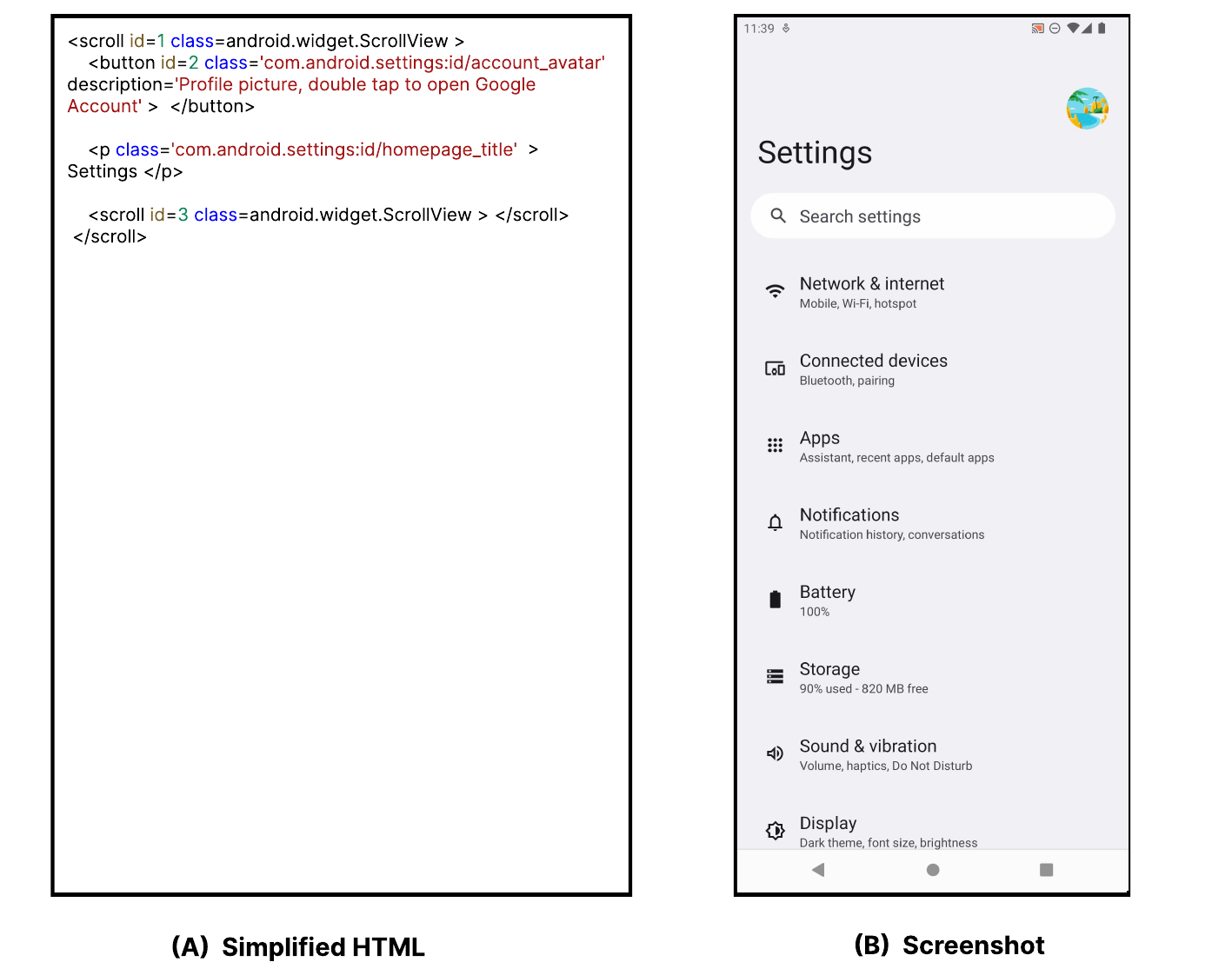}
  \caption{Example of incorrect UI extraction in an emulator. (A) The simplified HTML output shows only limited elements, missing most functions. (B) The corresponding screenshot of the Settings app displays the full range of options such as Network, Apps, Notifications, Battery, Storage, and Display.}
  \Description{Two panels compare emulator outputs. (A) Shows simplified HTML with only a few extracted elements: a profile picture button and the Settings title, while most UI components are missing. (B) Shows the actual screenshot of the Android Settings app with multiple categories including Network & internet, Connected devices, Apps, Notifications, Battery, Storage, Sound & vibration, and Display.}
\end{figure}

\section{Appendix C: Experiment Setup and Reproducibility} 
\subsection{Experiment Setup}
Following the experiment setup from Autotask\footnote{\url{https://github.com/BowenBryanWang/AutoTask}}, we developed and supported screenshot capture and multimodal prompting on top of it.

AutoTask natively supports a text-based UI representation (“screentext” derived from the Android UI hierarchy). We extend AutoTask with:

\begin{itemize}
    \item Screenshot acquisition: at each agent step, the system captures a screenshot of the current emulator screen and stores it alongside the screentext snapshot and step metadata.
    \item Multimodal prompt wiring:
        \begin{itemize}
        \item Text-only condition: prompt includes only screentext (structured UI representation).
        \item Multimodal condition: prompt includes screentext plus the screenshot captured at the same step.
    \end{itemize}
    \item Logging: we log, per step, the input screentext, screenshot path/identifier (multimodal only), predicted action, executed action, execution result, and reflection output.
\end{itemize}

\subsection{Prompt Snippet}
The following prompt was added to each agent in our modified agent system.
\begin{lstlisting}[style=cli, caption={Multimodal prompt instruction (with screenshot).}, label={lst:prompt-screenshot}]
A screenshot of the current UI is provided. Use it to ground your actions in the visible interface and to resolve ambiguities in the screentext (e.g., icons, visual grouping, or truncation).
\end{lstlisting}

\subsection{Code}
To support reproducibility, we will publicly release the complete evaluation pipeline (code, configs, and task files) upon publication, along with documentation to rerun the experiments. Here are some examples of this study.

\begin{lstlisting}[style=cli, caption={Example commands for running AutoTask with and without screenshots.}, label={lst:autotask-cli}]
# Without screenshot
python main.py --task "[YOUR_TASK]"

# With screenshot
python main.py --task "[YOUR_TASK]" --screenshot enable

# Examples
python main.py --task "Set an alarm for 6:30 AM"
python main.py --task "Set an alarm for 6:30 AM" --screenshot enable

python main.py --task "Open the file manager and navigate to the 'Documents' folder."
python main.py --task "Open the file manager and navigate to the 'Documents' folder." --screenshot enable

python main.py --task "On Facebook, check your latest post, like it and leave a comment, 'what a beautiful place', and share with your Feed and friend with 'My recent travel'."
python main.py --task "On Facebook, check your latest post, like it and leave a comment, 'what a beautiful place', and share with your Feed and friend with 'My recent travel'." --screenshot enable
\end{lstlisting}

\end{document}